\documentclass[%
reprint,
amsmath,amssymb,
aps,
prl,
]{revtex4-1}
\usepackage{graphicx}
\usepackage{amsthm,amsmath}
\usepackage[table,xcdraw]{xcolor}
\usepackage[top=25.4mm, bottom=20.9mm, left=19.1mm, right=19.1mm]{geometry}
\usepackage{mathtools}

\theoremstyle{definition} 

\newcommand{\DD}{\, \displaystyle}

\DeclareMathOperator{\EX}{\mathbb{E}}

\newcommand{\fracc}[2]{\, \displaystyle \frac{ #1}{ #2}}

\newcommand{\morabba}[1]{\,\begin{flushright}
\Rectsteel \\
\end{flushright}}

\newcommand{\all}[2]{\,\begin{align}
#1 
\label{#2}
\end{align}
}

\makeatletter
\newcommand{\vast}{\bBigg@{4}}
\newcommand{\Vast}{\bBigg@{5}}

\makeatother

\begin{document}
\preprint{APS/123-QED}
\title{ 
Evolution of Cooperation on Stochastic Block Models
}

\author{ Babak Fotouhi$^{1,2}$,  Naghmeh Momeni$^{1,3}$, Benjamin Allen$^{1,4,5} $, Martin A. Nowak$^{1,6,7} $\\
\emph{$^1$ Program for Evolutionary Dynamics, Harvard University, Cambridge, MA, USA}\\
\emph{$^2$ Institute for Quantitative Social Sciences, Harvard University, Cambridge, MA, USA}\\
\emph{$^3$  Massachusetts Institute of Technology (MIT) - Sloan School of Management, Cambridge, MA, USA}\\
\emph{$^4$ Department of Mathematics,  Emmanuel College, Boston, MA, USA}\\
\emph{$^5$ Center for Mathematical Sciences and Applications, Harvard University, Cambridge, MA, USA}\\
\emph{$^6$ Department of Mathematics, Harvard University, Cambridge, MA, USA}\\
\emph{$^7$ Department of Organismic and Evolutionary Biology, Harvard University, Cambridge, MA, USA} 
}

 \begin{abstract}  
 Cooperation is a major factor in the evolution of human societies. 
 The structure of human social networks, which affects the dynamics of cooperation and other interpersonal phenomena, have common structural signatures. 
 One of these   signatures is the tendency to organize as groups.
 Among the generative models that network theorists use to emulate this  feature is the  Stochastic Block Model (SBM). 
 In this paper, we study   evolutionary game dynamics on SBM networks. 
Using a recently-discovered duality between evolutionary games and coalescing random walks, we
  obtain analytical conditions such that natural selection favors  cooperation over defection. 
 We calculate the transition point for each community to favor cooperation. 
  We find that a critical inter-community link  creation probability exists  for given group density, such that the overall network supports cooperation even if individual communities inhibit it.
 As a byproduct, we  present mean-field solutions for the critical benefit-to-cost ratio which performs with remarkable accuracy for diverse generative network models, including those with community structure and heavy-tailed degree distributions. We also demonstrate the  generalizability  of the results to arbitrary two-player games.
 \end{abstract}

\maketitle
Cooperation is a central tenet of social life of many species.
 Cooperative dynamics are affected by the structure of social networks they take place on~\cite{lieberman2005evolutionary,szabo2007evolutionary,perc2013evolutionary,perc2010coevolutionary}. 
 A ubiquitous feature of social networks, including human social networks, is community structure, which involves the tendency of nodes to have different intergroup and intragroup connectivity~\cite{girvan2002community}. 
Previous studies have demonstrated the effects of such group structures. 
 A common model to generate and study community structure and to perform community detection is the Stochastic Block Model (SBM)~\cite{holland1983stochastic,decelle2011asymptotic,karrer2011stochastic}. 
 Here we  study the evolutionary dynamics of arbitrary symmetric 2$\times$2 games on SBM networks.
  We find analytical results such that natural selection would favor one strategy over the other. 
  We introduce a mean-field approximation that produces remarkably-accurate solutions which perform reasonably well even when the network size is as small as 20. 
 We utilize the solutions to study how inter-connection of seggregated communities under the SBM framework affects the fate of cooperation in the whole network. 
 We highlight that our results give  analytical support to the previous   studies in the literature which demonstrated the effect of inter-group connectivity   on the evolution of cooperation with diverse simulation setups and model settings~\cite{wang2013optimal,wang2013interdependent,jiang2013spreading,wang2014rewarding}. Together, these findings evince a robust phenomenon regarding how  inter-group connectivity affects collective cooperation. 
 We conclude by demonstrating that the proposed solutions   perform remarkably well for network models other than SBM, including those with heavy-tailed degree distributions.

 We consider a generic $2\times 2$ game with two available strategies, C and D.  Both players receive $R$ if they mutually play  C, and 
receive $P$  if they mutually play D. 
If one player plays C and the other plays D, the C-player receives $S$ and the  D-player receives $T$. 
The strategy  of node $x$ is denoted by $s_x  \in \{0,1 \}$, where 1 corresponds to strategy C and 0 to strategy D. 
We denote the set of network neighbors of node $x$ by $    \mathcal{N}_x$. 
We consider averaged payoffs, where the payoff of node $x$   is given by
 \all{
 \resizebox{.98\linewidth}{!}{$
 f_x=\DD  \sum_{y    \in \mathcal{N}_x}   s_x \big[R s_y + S (1-s_y)\big] +  (1-s_x) \big[T s_y + P (1-s_y)\big]
 .
 $}
 }{fx}
At each timestep, a random  individual is chosen to update its strategy via copying that of a neighbor. 
 The probability that neighbor  $x$ is copied is proportional to $1+\delta f_x$, where $f_x$ is the payoff of $x$ and ${0<\delta \ll 1}$ models the selection strength, that is, higher $\delta$  indicates stronger social learning via observing the payoff  of peers. The case $\delta=0$  
is equivalent to  the \emph{voter model}~\cite{sood2005voter,antal2006evolutionary,sood2008voter}. The weak-selection limit considered here can be viewed as the first-order correction to the voter model. 
  We seek  an analytical condition such that natural selection favors the fixation of C over fixation of  D.
   According to the Structure Coefficient Theorem~\cite{tarnita2009strategy}, this happens if
  ${  (R-P)\sigma >(T-S)}$, where $\sigma $ is the structure coefficient, which is independent of the game and only depends on the network structure.
 Thus, it suffices to calculate $\sigma$ for the given network. 
 This can be done \emph{exactly} by considering a mathematical  equivalence between the evolutionary game dynamics and that of coalescing random walks~\cite{allen2017evolutionary}. Below we briefly outline this recent exact framework, highlight the computational cost of the solution for very large networks, propose a  mean-field approximation to ameliorate the situation for large networks, and then apply this mean-field solution to obtain remarkably accurate solutions for SBM networks. After analyzing the case of SBM networks in detail, we conclude by pointing out the reasonable performance of the proposed mean-field solution for a wide array of other network models.  
 
  To obtain $\sigma$ for generic $2\times2$ games, we apply the methodology of~\cite{allen2017evolutionary} to unweighted undirected graphs. 
  We analyze  the `donation game' version of the Prisoner's Dilemma, with ${R=b-1}$, ${S=-1}$, ${T=b}$, and ${P=0}$ and later discuss how $\sigma$ can be readily  derived from the results.  
  The payoff of node $x$ is given by: 
  \all{
  f_x(t) = -c s_x(t)  + \frac{1}{k_x} \sum_{y    \in \mathcal{N}_x} b s_y(t)
  .}{fxpd}
 
 For the update of node $x$, the following holds: 
 \all{
\EX{[s_x(t+1)]}
&=  (1-\frac{1}{N}) s_x(t)
 \nonumber \\ &
 + \fracc{1}{N} \sum_{y    \in \mathcal{N}_x} \fracc{1+ \delta f_y(t)}{\sum_{z   \in \mathcal{N}_x} [1+ \delta f_z(t)]} s_y(t) 
.}{sx1}
In the limit of weak selection, expanding  to the first order of $\delta$ after multiplying both sides by $k_x$, we get:
\all{
&
\EX{[k_x s_x(t+1)]} = 
  \fracc{1}{N }  \bigg[ \sum_{y    \in \mathcal{N}_x} s_y(t) +\delta \sum_{y    \in \mathcal{N}_x} f_y(t) s_y(t) 
  \nonumber \\ &
  -\sum_{y,z    \in \mathcal{N}_x  }\delta  \frac{ s_y(t) f_z(t)}{k_x}  \bigg]+ O(\delta^2)
  +(1-\frac{1}{N}) k_x s_x(t)
 .}{sxt2}
 Now we define $\psi(t) := \sum_x k_x s_x(t)$.
 Summing~\eqref{sxt2} over all nodes, we see that in the zeroth-order dynamics of the system (i.e., $\delta=0$, corresponding to the voter model),   the expected value of $\psi(t)$ is a conserved quantity~\cite{sood2005voter,sood2008voter,fotouhi2014voter}. 
 The fixation probability can be obtained by equating the average value of $\psi$ over every initial mutant placement with its expected value as $t \rightarrow \infty$. 
%
 The expected first-order change of $\psi$  is 
 \all{
 &
 \Delta \psi^{(1)} (\vec{s}) =  
   \fracc{\delta}{N }  \bigg[  \sum_x \sum_{ y \in \mathcal{N}_x} f_y  s_y 
  - \sum_x \sum_{y,z   \in \mathcal{N}_x}\frac{  s_y f_z }{k_x}   \bigg] 
  \nonumber \\ &
  =  \fracc{\delta}{N }  \bigg[  \sum_x k_x f_x s_x
  - \sum_x \sum_{y,z   \in \mathcal{N}_x}\frac{  s_y f_z }{k_x}   \bigg] 
    \nonumber \\ &
  =  \fracc{\delta}{N } \sum_x  \bigg[  k_x   s_x \Big(-c s_x +\frac{b}{k_x} \sum_{ y \in \mathcal{N}_x} s_y \Big)
      \nonumber \\ &
  - \sum_{y,z   \in \mathcal{N}_x}\frac{  s_y  \Big(-c s_z +\frac{b}{k_z} \sum_{w   \in \mathcal{N}_z} s_w \Big) }{k_x}   \bigg] 
      \nonumber \\ &
}{deltapsi}
Exchanging the summation order,   denoting the first moment of the degree distribution (the average degree) by $\mu_1$, we get
\all{
 \Delta \psi^{(1)} (\vec{s})&=  
  -\fracc{\delta}{N}  c (N\mu_1)  +   \fracc{\delta}{N } \sum_{y}  \bigg[      b \sum_{x \in \mathcal{N}_y } s_y    s_x 
      \nonumber \\ &
  +c \sum_{x  \in \mathcal{N}_y} \sum_{z   \in \mathcal{N}_x} \frac{    s_y   s_z}{k_x} 
  - b\sum_{x   \in \mathcal{N}_y} \sum_{z   \in \mathcal{N}_x} \sum_{w   \in \mathcal{N}_z}  \frac{  s_y s_w   }{k_x k_z}   \bigg] 
  .
 }{deltapsi2}
 We need to sum up these expected increments from $t=0$ up to $t=\infty$. Let $\xi_x$   denote this expected total change for given initial condition in which only node $x$ is C and all other nodes are D . The fixation probability for a given initial condition will then be $(k_x+\xi_x)/(N \mu_1)$. Averaging over all nodes, this becomes: 
 \all{
 \rho= \frac{1}{N} + \fracc{1}{N} \sum_x \xi_x.
 }{rhoxi}
 The sum on the right hand side of~\eqref{rhoxi}  requires the calculation of temporal sum of  the  spin products on the right hand side of~\eqref{deltapsi2}. 
 Note that these summations are to be performed  in the voter-model regime. That is, due to the factor $\delta$, we should only keep the summations in zeroth order. 
  In~\cite{chen2013sharp,allen2017evolutionary}, it is shown that for any two nodes $i$ and $j$,  if we find the expected  temporal sum of  $1/N-s_i s_j$   from $t=0$ to $t=\infty$ under the voter-model dynamics and then average this sum over all single-node initial placements,   the result is equal to ${ \tau_{ij}/(2N)}$, where $\tau_{ij}$ is the expected meeting time of two random walkers initiated at nodes $i$ and $j$. 
These meeting times follow the following recurrence relation:
\all{
\tau_{ij}= \tau_{ji} = 
(1-\delta_{ij} )
\Bigg[ 
1+\frac{1}{2 k_i} \sum_{\ell \in \mathcal{N}_i}   \tau_{\ell j}
+\frac{1}{2 k_j} \sum_{\ell\in \mathcal{N}_j}  \tau_{\ell i}
    \Bigg] 
.}{sys}
Thus the random walk equivalence  relates the fixation probability  to the expected value of the meeting times of two random walkers initiated one, two, and three steps away on the network, corresponding to the three last terms on the right hand side of Equation~\eqref{deltapsi2}, respectively. 
Note that  the expression `$\ell$ steps away' here  refers to random-walk steps,  rather than graph distance. 
So for example,  node $y$ is $\ell$ steps away from node $x$ if $A^{\ell}_{xy}>0$, where $A^{\ell}$ is the $\ell$-th power of the adjacency matrix.
 Using the meeting times which are the solutions to the system of equations~\eqref{sys}, we define the   quantity $\tau_x$ as  the expected remeeting time of two random walkers both initiated at node $x$: 
 \all{
 \tau_x  = 1+\fracc{1}{k_x}\sum_{y\in \mathcal{N}_x}   \tau_{yx}
 .}{taux}
 We also define $p_x=\sum_{y   \in \mathcal{N}_x} 1/(k_xk_y)$. 
 Using these definitions, and after some algebraic simplifications,  the fixation probability can be expressed in the following   form: 
\all{
\rho=\fracc{1}{N}+ \fracc{\delta }{2N} \bigg[ & b \big(\sum_x \fracc{k_x}{N \mu_1} \tau_x -2\big) 
\nonumber \\ &
-c \big(\sum_x \fracc{k_x}{N \mu_1} \tau_x p_x-2 \big)\bigg]  + O(\delta^2)
.
}{fix}
The critical benefit-to-cost ratio for C to be favored by natural selection is thus given by setting this fixation probability be greater than neutral drift, which yields: 
\all{
b^*  = \fracc{\sum_x \tau_x k_x  - 2 N \mu_1}{\sum_x \tau_x k_x p_x - 2 N \mu_1}
.}{gamma}

The drawback of this exact framework is that solving~\eqref{sys} requires solving a system of $N(N-1)/2$  linear equations, which can be infeasible for very large networks. The conventional Cholesky decomposition methods to solve this system have the complexity of order $N^6$, and  faster techniques are mainly  applicable only when the network is sparse. 
So for large   networks in general, which can as well be dense, obtaining the exact solution is computationally infeasible. 
To obviate this limitation, here we seek a mean-field approximation to obtain   analytical  results that are computationally feasible for large networks. 
We use the fact that, combining~\eqref{taux} and~\eqref{sys}, the remeeting times of these random walkers satisfy the following equation: 
\all{
\sum_x k_x^2 \tau_x = N^2 \mu_1^2 
.}{tau_iden}
Using this equation and assuming a mean-field approximation in which every $\tau_x$ value is replaced by the average value over all $x$, we get: 
\all{
\tau_x=N \fracc{\mu_1^2}{\mu_2} 
,}{tauxMF}
where $\mu_2$ is the second moment of the degree distribution. 
For the mean-field approximation of the fixation probability, we plug this into~\eqref{fix} and obtain:
\all{
\rho \approx \fracc{1}{N}+ \fracc{\delta }{2N} \bigg[   b \big(N \fracc{\mu_1^2}{\mu_2}  -2\big) 
-c \big( \fracc{\mu_1}{\mu_2} \sum_x k_x p_x-2 \big)\bigg]   
}{fixmf1}
Noting that $\sum_x k_x p_x= \sum_x \sum_{y\in N_x} 1/k_y$ is equal to $N$ for any network, we get: 
\all{
\rho \approx \fracc{1}{N}+ \fracc{\delta }{2N} \bigg[  b \big(N \fracc{\mu_1^2}{\mu_2}  -2\big) 
-c \big( \fracc{N\mu_1}{\mu_2} -2 \big)\bigg]   
}{fixmf2}
Plugging this result into~\eqref{gamma}, we  arrive at: 
\all{
b^*  \approx \fracc{N - 2 \fracc{\mu_2}{\mu_1^2}}{\fracc{N}{\mu_1} - 2 \fracc{\mu_2}{\mu_1^2}}
}{good}
 
\begin{figure}[t]
\centering
\includegraphics[width=.98 \columnwidth]{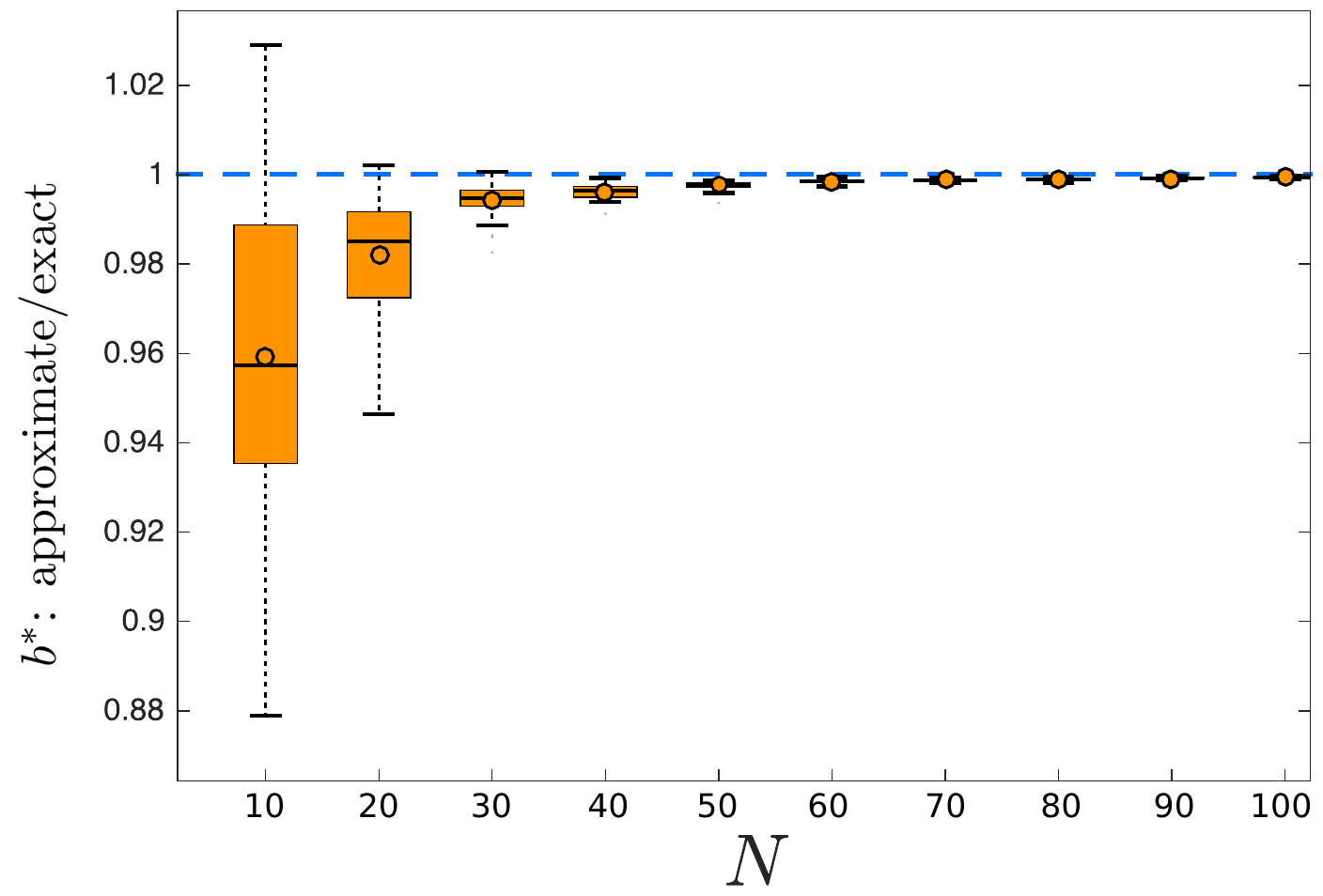}
\caption  
{
 \footnotesize
 (Color Online) 
Accuracy of the proposed mean-field  approximation for $b^*$ as a function of $N$. The network generation parameters are set to the example values of  $p=0.7$, $q=0.1$, and $m=3$. The performance is consistently well for every parameter configuration tried, which will be demonstrated in Figure~\ref{families}. 
  }\label{SBM_Ns}
\end{figure}

We consider a stochastic block model~\cite{holland1983stochastic,decelle2011asymptotic} with $m$ equi-probable groups, intra-community link probability $p_{}$ and inter-community link probability $q_{}$. 
  For   node $x$ with degree $k_x$,  denote the number of within-community neighbors by $k_x^{\text{intra}}$ and denote the number of its neighbors in other communities by $k_x^{\text{inter}}$. For large $N$, the average degree $\mu_1$, which is the expected value of $k_x$, is 
  \all{
  \resizebox{\linewidth}{!}{$
  \mu_1  = \EX{(k_x^{\text{intra}})} +\EX{(k_x^{\text{inter}})} = p_{} \fracc{N-1}{m}+   q_{} (N-1) \left( 1-\fracc{1}{m} \right)
  .
  $}
  }{mu1}
Also because the inter and intra-community degree distributions are independent, the variance of $k$ is the sum of the variance of $k_x^{\text{intra}}$ and the variance of $k_x^{\text{inter}}$: 
  \all{
\resizebox{\linewidth}{!}{$
 \text{var}(\text{k})
  = p_{}(1-p_{}) \fracc{N-1}{m}+   q_{}(1-q_{})(N-1)  \left( 1-\frac{1}{m} \right)
  .
  $}
  }{var}
Combining~\eqref{mu1} and~\eqref{var}, we obtain $\mu_2$, which we can insert into~\eqref{fixmf2} to obtain the fixation probability. 
For the critical benefit-to-cost ratio, we insert the expression for $\mu_2$ and $\mu_1$ into~\eqref{good}. 
Defining ${\alpha:=1/m}$ and ${\beta:=1-1/m}$ for brevity,   and after algebraic simplifications, we obtain:
\all{
\resizebox{\linewidth}{!}{$
b^* _{\text{SBM}}
\approx
 \fracc{N - 2 -\fracc{2}{N-1}\fracc{  \alpha p_{}(1-p_{}) +    \beta q_{}(1-q_{}) }{(\alpha p_{} +   q_{}   \beta  )^2}}{\fracc{N/(N-1)}{   \alpha p_{} +   \beta  q_{} } - 2-\frac{2}{N-1}\frac{  \alpha p_{}(1-p_{}) +    \beta q_{}(1-q_{}) }{(  \alpha p_{} +   \beta  q_{}  )^2}}
.
$}
}{goodSBM}
 Figure~\ref{SBM_Ns}  demonstrates that  the approximation~\eqref{goodSBM}  has relative error less than 1\% for network size as small as 40. 
 In these network sizes, the exact method can be employed in reasonable time and thereby we have a benchmark to assess the solutions. 
 Figure~\ref{SBM_Ns} demonstrates that for large networks, the error rate is remarkably small, so for large networks where the exact method become prohibitively constly

\begin{figure}[t]
\centering
\includegraphics[width=.98 \columnwidth]{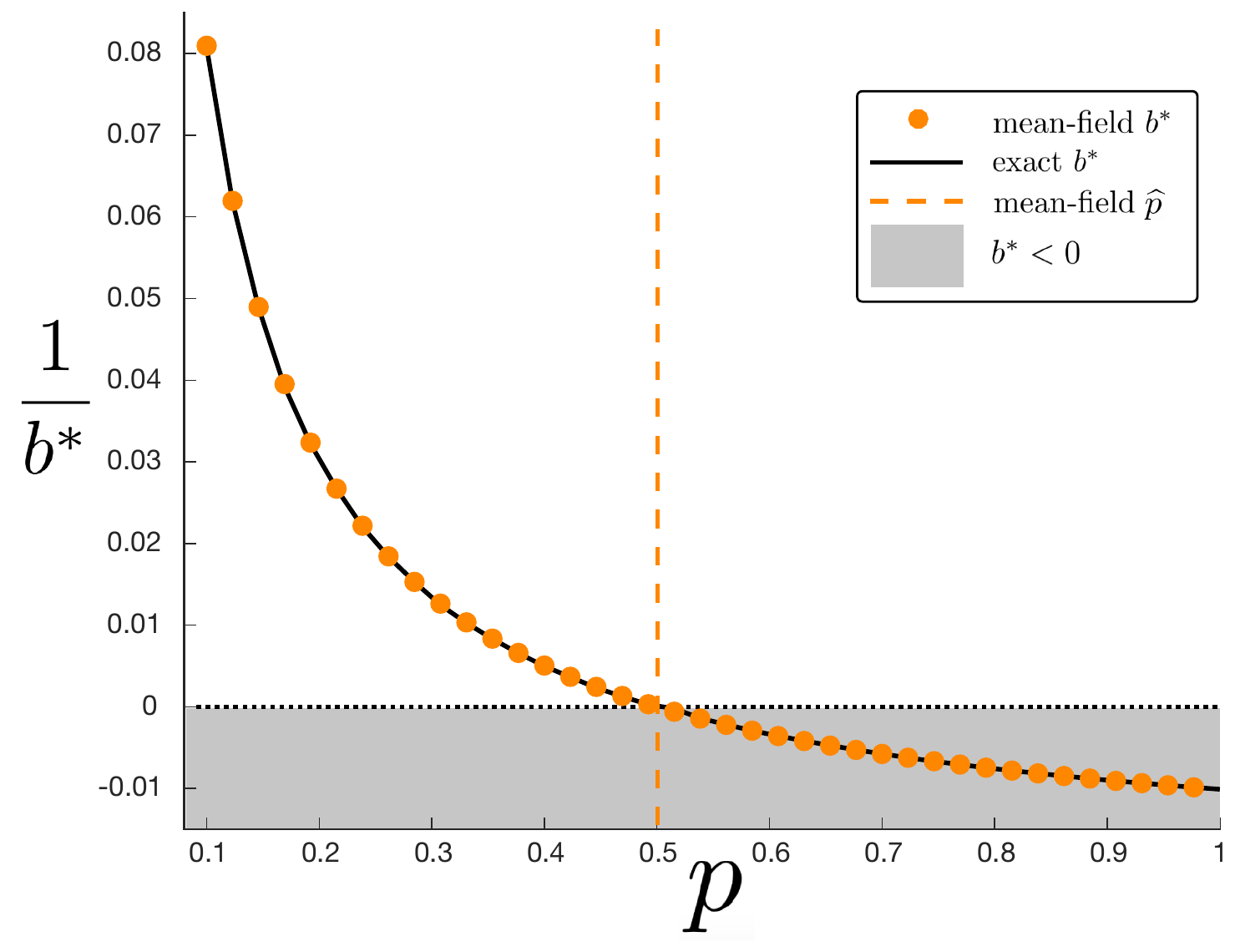}
\caption  
{
 \footnotesize
 (Color Online) 
Accuracy of the proposed mean-field  approximation for $b^*$ for ER networks as a function of link creation probability $p$.  
As predicted by Equation~\eqref{goodER}, there is a phase transition at $\widehat{p}=1/2$, above which natural selection does not favor cooperation over defection regardless of the benefit-to-cost ratio. 
The mean-field prediction    is in agreement with   the exact results. 
  }\label{ER}
\end{figure}

The special case of $p=q$ is equivalent to the Erd\H{o}s-R\'enyi  (ER) model~\cite{ER1}. Equivalently, we can set $m=1$. 
In this special case Equation~\eqref{goodSBM} simplifies to: 
\all{
b^* _{\text{ER}}
\approx
\fracc{p(N^2-3N+4)-2}{(N-2)(1-2p)}   
.}{goodER}
Figure~\ref{ER} illustrates the accuracy of the proposed mean-field  approximation~\eqref{goodER} for ER networks. 
We plot $1/b^*$ instead of $b^*$ because it gives visually better results. 
Equation~\eqref{goodER} indicates a phase transition at $\widehat{p}=1/2$, which is visible in Figure~\ref{ER}. 
That is, in the ER model, the  expected value  of the critical benefit-to-cost ratio becomes negative if $p>\widehat{p}$. 
In this regime, the fixation probability of a cooperative mutant is less than that of neutral drift, regardless of the values of $b$ and $c$. 
Hence natural selection does not favor the fixation of cooperation over the fixation of defection for \emph{any} value of benefit-to-cost ratio.
In this regime, the network promotes spite, which means that players are willing to pay a cost to reduce the payoff of others.
Hence we observe a phase transition from cooperation to spite governed by network density.

\begin{figure}[t]
\centering
\includegraphics[width=.98 \columnwidth]{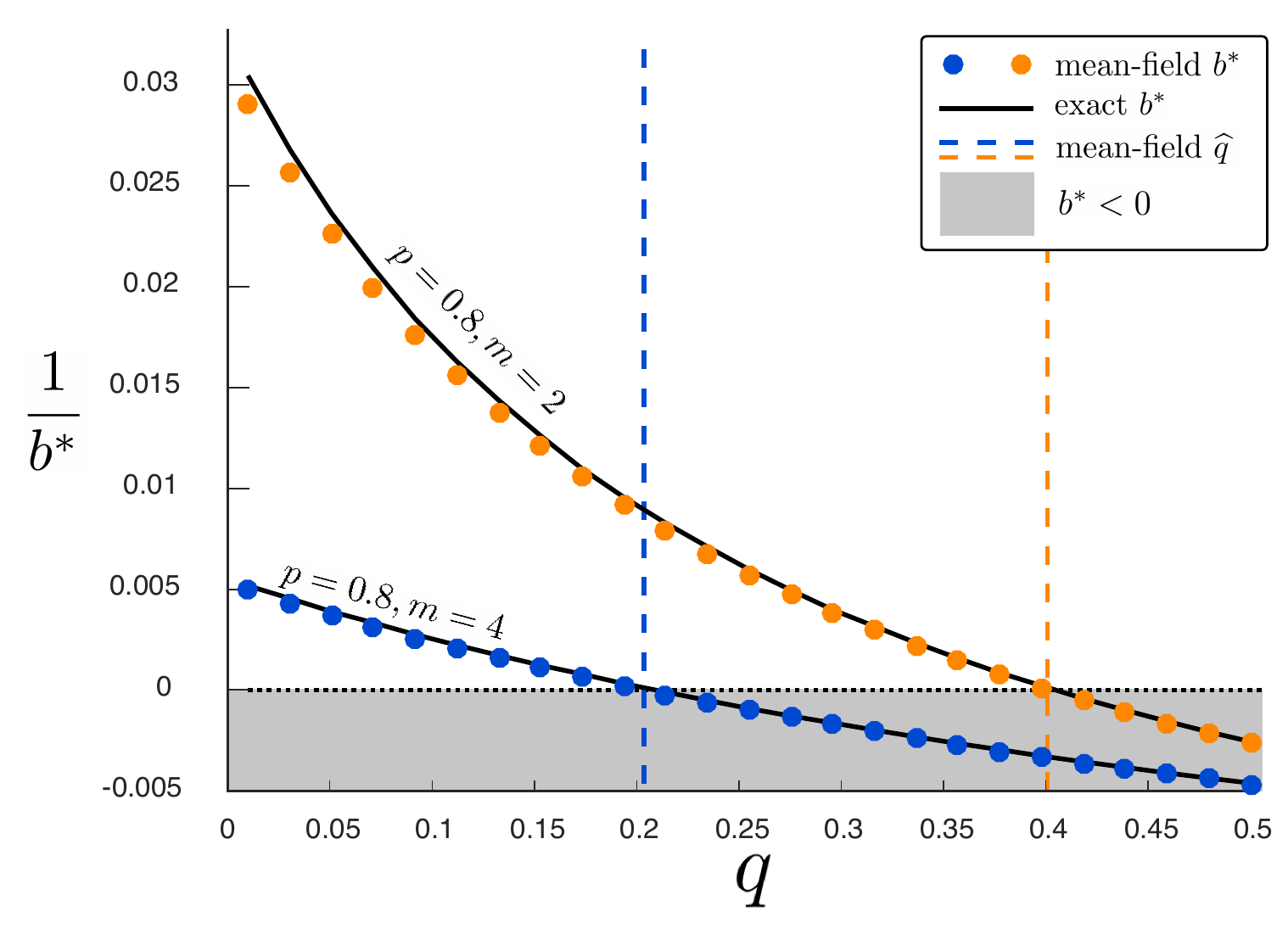}
\caption  
{
 \footnotesize
 (Color Online) 
Accuracy of the proposed mean-field approximation for $b^*$ for the example value of $p=0.8$, and two different values of $m$. 
The network size is 100. 
The dashed lines depict the predicted threshold value $\widehat{q}$  as given by~\eqref{SBM_phase}, agrees with the exact results for both settings. 
For $q>\widehat{q}$, natural selection promotes fixation of defection over cooperation regardless of the benefit-to-cost ratio. 
The intercepts match the result for $\lim q \rightarrow 0^+$ obtained in Equation~\eqref{smallq}. 
  }\label{SBM_qhat_good}
\end{figure}

For the general case of $p_{} \ne q_{}$, too, we can find such a point of transition. 
We can find  $\widehat{q }$ (the critical inter-community link probability above which natural selection does not favor fixation of cooperative mutants over defective mutants regardless of the benefit-to-cost ratio) by setting the denominator of Equation~\eqref{goodSBM} equal to zero and solving the resulting quadratic equation for $q$.  
Expanding the solution for large $N$, we get: 
\all{
\widehat{q} =\fracc{m-2p_{}}{2(m-1)} + 2m \fracc{\left(\frac{1}{2}-p_{}\right)^2}{(m-1)^2} \fracc{1}{N}+ O\left(\fracc{1}{N^2}\right)
.
}{SBM_phase}
This result has an  important consequence. 
Each of the $m$ communities with intra-community link probability $p_{}$ considered separately is an ER network, so the expected value of the critical benefit-to-cost ratio for each of them is given by~\eqref{goodER}. 
 Suppose the communities have $p_{}>1/2$, which, as discussed above,  means that cooperation is not favored by natural selection for each individual community considered separately. 
 We can then interconnect these communities under the SBM setting, with inter-community probability $q_{}$. 
For $q_{}<\widehat{q} $, the critical benefit-to-cost ratio of the overall network  is positive, despite individual communities inhibiting the fixation of cooperation. This confirms analytically  the numerical observations about conjoining random networks~\cite{conjoining}. 

In Figure~\ref{SBM_qhat_good}, we present the comparison of Equation~\eqref{goodSBM} with the exact results. The orange markers pertain to the example case of $m=4$ and $ p=0.8$, for which Equation~\eqref{SBM_phase} gives $\widehat{q} \approx 0.4$.
 The blue markers pertain to $m=2$ and $ p=0.8$, with $\widehat{q} \approx 0.2$. 
 The network size is 100 in both cases. 
 The approximations are   remarkably close to the exact values.

Of particular relevance  for actual scenarios is the case where $q \ll 1$, which means that the communities are sparsely interconnected. In this regime, we can  expand  $b^*$   as follows: 
\all{
b^*
&= 
\fracc{N(N-2)p-2m(1-p)}{N(m-2p)-2m(1-p)}
+ O(q)
.}{smallq}
The interesting result here is the existence of the zeroth-order term. 
 This fact can be seen in Figure~\ref{SBM_qhat_good} as well. 
 For $N=100,p=0.8$ with $m=2$, from Equation~\eqref{smallq} we get $b^*  \approx 200$, and with $m=4$ we get $b^*  \approx 32.9$. 
 The  inverse of these values are $0.0050$ and $0.030$, respectively, which correctly  match the intercepts observed  in Figure~\ref{SBM_qhat_good}. 
 This confirms that in the limit as $q \rightarrow 0^+$, $b^*$ tends to a  positive number. 
 Thus,  sparse interconnection of cohesive communities rescues cooperation. 
 Note that this happens as long as  the whole network is connected, so that $b^*$ is well-defined. 
 The minimum value of $q$ such that the whole network is connected goes to zero as $N$ tends to infinity. 

\begin{figure}[t]
\centering
\includegraphics[width=.98 \columnwidth]{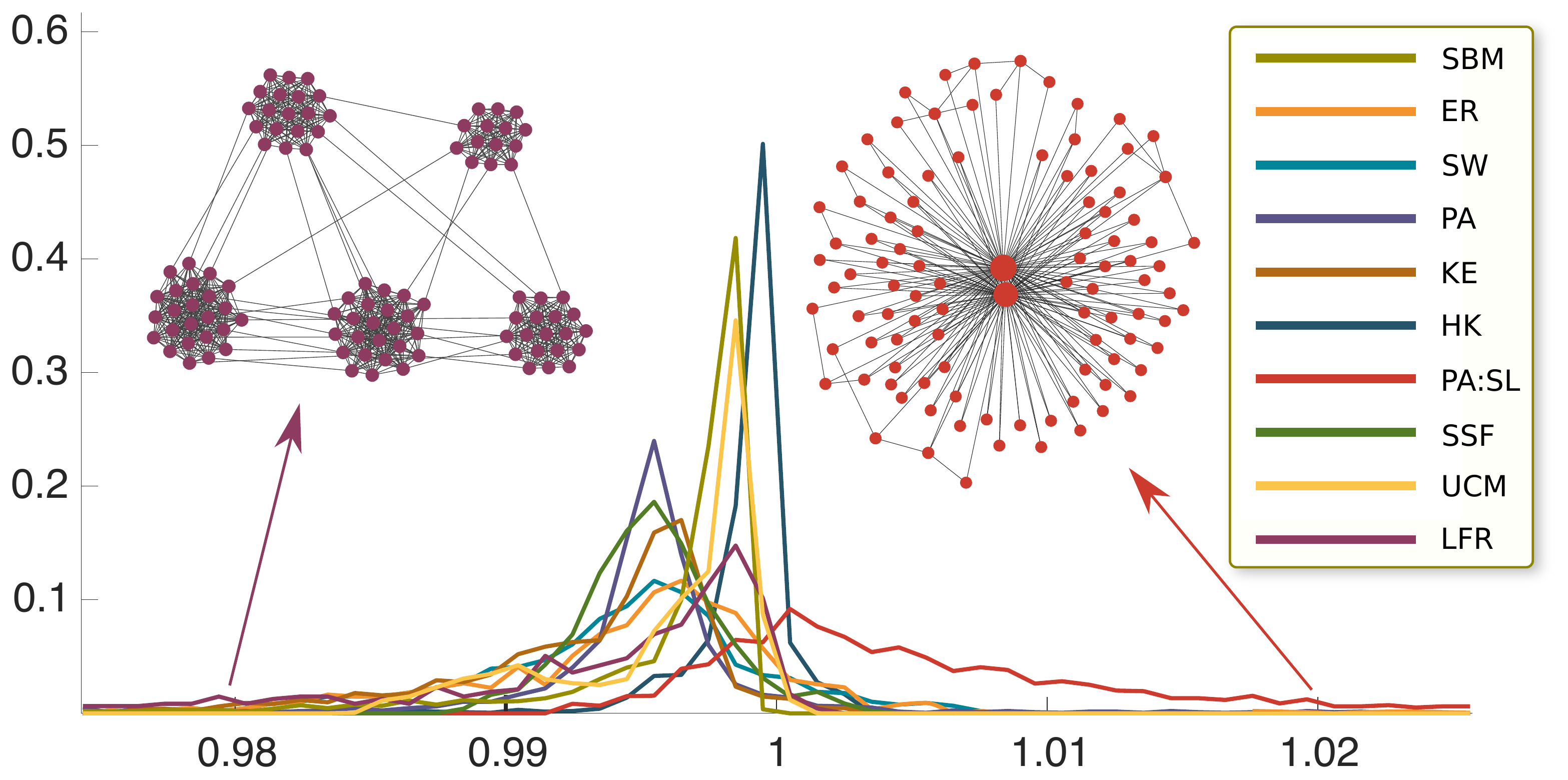}
\caption  
{
 \footnotesize
 (Color Online) 
The distribution of the ratio of the proposed  mean-field approximation to the true value of $b^*$. 
The true value is obtained by solving the $N(N-1)/2$ linear equations as given by~\eqref{sys}, which is infeasible for very large networks, but feasible for the size ranges of the test networks. The 10 network generation families are discussed in the text. 
Two example graphs with high heterogeneity are depicted to highlight the robustness of the proposed mean-field approximation to   structural heterogeneity.
  }\label{families}
\end{figure}

%
%

Returning to arbitrary $2\times 2$ games, according to the Structure Coefficient Theorem, the structure coefficient is defined in terms of $b^*$ as follows: 
\all{
\sigma =\fracc{b^*+1}{b^*-1}
.}{sigma0}
 For the special case of SBM or ER networks, $\sigma$ can be obtained readily by inserting the corresponding values of $b^*$  from Equations~\eqref{goodSBM} and~\eqref{goodER}   into~\eqref{sigma0}, respectively.  
 For general networks, we use the   mean-field value for $b^*$ obtained in~\eqref{good}. The result simplifies to the following:
\all{
\sigma  
  \approx
 \fracc{N(\mu_1+1)-4 \mu_2/\mu_1}{N(\mu_1-1)} 
.}{sigma}
Natural selection favors the fixation of strategy C over the fixation of strategy D if   ${  (R-P)\sigma >(T-S)}$.
  Thus we have obtained analytical conditions such that natural selection favors the fixation of one strategy over the other   for any $2\times 2$ game. 

We conclude by highlighting  the accuracy of the proposed mean-field approximation for $b^*$ via simulations and numerical results for various graph families, in addition to the SBM and ER models. 
In Figure~\ref{families}, we plot the histogram of the ratio of the approximate $b^*$ to the exact value (computed via solving the system of $N(N-1)/2$ linear equations as given by~\eqref{sys}) for 10 different network families. 
In addition to SBM and ER networks, we consider the Small-world model, and 7  other network generation families, which all generate networks with heavy-tailed degree distribution. 
We use these additional 7 families as pessimistic scenarios regarding our mean-field approximation, because these models generate highly-heterogeneous networks. 
For the SBM model, we chose $m$ uniformly from $\{2,3,4,5\}$, chose $p_{}$ uniformly in $[0.1,1]$, and $q_{}$ uniformly in $[.01,p]$. 
For the small-world model~\cite{newman1999renormalization}, we chose the initial lattice degree uniformly from $ \times \{4,8,12\}$ and the link creation probability uniformly from $[0,0.1]$. 
For ER networks, we chose the link formation probability uniformly in $[0.2,1]$. 
For preferential attachment (PA) with shifted-linear kernel~\cite{dorogovtsev2000structure} (also called `initial attractiveness'), we randomly generate $m$ between 1 and 5, and the kernel bias is generated randomly between 0 and 5 (the closer to zero, the closer the model is to the Barabasi-Albert model). 
For the scale-free model of Holme and Kim  (HK)~\cite{holme2002growing}, we chose the triad formation probability of the model uniformly in $[0,1]$. 
For  Klemm-Euigluz   (KE) scale-free model~\cite{klemm2002growing}, we choose the cross-over probability parameter of the model uniformly in $[0,1]$. 
For both models, we choose the number of initial connections of incoming nodes uniformly between 1 and 5.
For the spatial scale-free model of Barthelemy~\cite{barthelemy2003crossover} (SSL), we generated networks on a 2D lattice with distance decay parameter $r_c$ chosen uniformly in $[0,0.2]$. 
For the uncorrelated configuration model (UCM)~\cite{catanzaro2005generation}, we chose the minimum number of connections uniformly between 1 and 5, and the exponent in the power-law degree distribution is chosen uniformly in $[1,4]$. 
For super-linear preferential attachment~\cite{krapivsky2000connectivity} (PA:SL), where the attachment kernel depends on degrees as $k^\theta$, we chose the number of initial connections of incoming nodes uniformly between 1 and 4, and the exponent of the kernel uniformly in $[0,3]$. Greater exponents produce networks with higher degree inequality. 
The LFR benchmarks~\cite{lancichinetti2008benchmark} are generated with the mixing parameter   chosen uniformly at random in  $[0,1]$. The maximum degree    $k_{\text{max}}$ was chosen uniformly between 1 and $N-1$, the average degree was chosen uniformly between 1 and $k_{\text{max}}$, the degree exponent was uniformly in $[0,3]$, the exponent for the community size distribution was chosen uniformly in  $[0,1]$.
For each family, we generate 10000 networks. 
For every generate network, the size is randomly chosen between 100 and 500, which is reasonably small so that the exact results could be feasibily calculated via solving Equation~\eqref{sys}. 
The histograms are highly concentrated around unity, which confirms the accuracy of the proposed mean-field approximations. 
Interestingly, the proposed approximation works well for networks with high structural heterogeneity, including those with heavy-tailed degree distributions.

%

We  presented accurate mean-field solutions for    $2\times2$ games on heterogeneous networks that  determine which strategy is favored by natural selection. 
We have utilized our solution to study the case of  SBM networks in detail, and have uncovered a network-structural phase transition which pertains to regions in which one of the strategies will not be favored by natural selection regardless of the payoff parameters. 
 We have obtained similar results for ER networks as a special case. 
  We have obtained analytical expressions for the inter-connection of  segregated communities, which individually inhibit cooperation, but after interconnection, can collectively favor cooperation. This result in agreement with the previous results in the literature~\cite{jiang2013spreading,wang2013optimal}, which  obtain qualitatively similar results via various updating schemes and simulation parameters, and even those that consider many-player games~\cite{wang2013interdependent}.  
   Our analytical  findings  together with  the  previous simulation studies   consistently   highlight  the  cooperative advantage of sparsely interconnecting cohesive communities, and indicate that this advantage  is a robust feature of cooperative dynamics on networks, which has notable real-world consequences.

%


%

\end{document}